\documentclass[prl,preprint,showpacs]{revtex4}
\usepackage{amsmath,amsfonts}
\usepackage{epsfig}

\newcommand{\la}{\lambda}
\newcommand{\bea}{\begin{eqnarray}}
\newcommand{\beq}{\begin{equation}}
\newcommand{\eea}{\end{eqnarray}}
\newcommand{\eeq}{\end{equation}}
\def\simgeq{\; \raisebox{-0.4ex}{\tiny$\stackrel
{{\textstyle>}}{\sim}$}\;}

\begin{document}
\title
{On the Thermodynamic Limit of the Lipkin Model}
\date{\today}
\author{W.D.~Heiss}

\affiliation{Institute of Theoretical Physics and Department of Physics,
University of Stellenbosch, 7602 Matieland, South Africa }

\begin{abstract}
The thermodynamic limit of the Lipkin model is investigated. While the limit turns
out to be rather elusive, the analysis gives strong indications that the limit
yields two analytically dissociated operators, one for the normal and one for the 
deformed phase. While the Lipkin Hamiltonian is hermitian and has a second order
phase transition in finite dimensions (finite particle number), both properties
seem to be destroyed in the thermodynamic limit. 
\end{abstract}

\maketitle

\section{Introduction} The Lipkin model \cite{lip}, even though it appeared more 
than forty years ago in the literature, has, up to present, given rise to a vast 
amount of research papers. This continuing interest is due to its basic properties in 
that it demonstrates the mechanism of a phase transition including symmetry breaking 
for a schematic many body system. Originally introduced as a model in nuclear physics 
\cite{ring} it has found applications in a broad 
range of other topics: statistical mechanics of quantum spin systems \cite{bot82},
Bose--Einstein condensates \cite{cir98} as well as quantum entanglement \cite{vid04}, 
to name but a few. Recent interest is focused upon its large $N$ behaviour with 
$N$ being the particle number. Progress has been achieved using a variety of methods 
such as continuous unitary transformations \cite{sch}, a semiclassical 
approach \cite{ley} or the Bethe ansatz \cite{dus}. 

We briefly list here the major results relevant to the present paper.
In its original form the Lipkin model considers interacting Fermions
occupying two $\Omega$-fold degenerate levels. One major appeal is due to its easy
solubility. This is achieved by rewriting the kinetic energy term of the
Fermion Hamiltonian, {\it viz.}
$$\sum _{k,\sigma}\sigma a^{\dagger }_{k,\sigma} a_{k,\sigma} $$
and the two body interaction term
$$\sum_{k,k',\sigma}a^{\dagger }_{k,\sigma}a^{\dagger }_{k',\sigma}a_{k',-\sigma}a_{k,-\sigma} $$
in terms of the SU(2) operators $J_z$ and $J_{\pm }^2$, respectively.
In dimensionless form the Hamiltonian thus reduces to
\beq
H(\la )=J_z+\frac{\la}{2N}(J_+^2+J_-^2)
\label{lip}
\eeq
where the $2j+1=N+1$-dimensional representations of the SU(2) operators are used with
$N\le \Omega$ being the particle number.
Here the interaction is scaled by $N$ to ensure that $H$ is extensive, the operators 
$J_+^2$ and  $J_-^2$ effectively scale as $N^2$.
The Hamiltonian allows reduction into two spaces: 
$m$ integer and $m$ half-integer, with $m$ the eigenvalues of $J_z$;
it corresponds to $N$ even and odd, respectively and is denoted as parity.
For $\la \simgeq 0$ the $N$-even and $N$-odd levels are obviously separated and remain 
so for all $\la <1$ while the levels become degenerate for $\la >1$. The phase at 
$\la <1$ is called the normal phase while the symmetry (parity) breaking phase at 
$\la >1$ is called the deformed phase. For $\la \simgeq 1$ this phase transition is 
confined to the lower part of the spectrum in that higher up in the spectrum the normal 
phase still prevails. In fact, for a specific $\la  >1$ there is an energy 
$E_{k_c}(\la )$ such that the ordered spectrum $E_k$ is for $k<k_c$  associated with 
the deformed phase while for $k>k_c$ the energies relate to the normal phase \cite{hescge}.
In other words, the transition moves up in the spectrum from the ground state $E_1$ to
$E_{N/2}$ when $\la $ increases from unity to infinity (we confine ourselves to the lower
half of the spectrum as the symmetry of the model yields the same pattern in the 
upper half but  mirror reflected). For finite $N$, $E_{k_c}$ is clearly characterised by
the minimum of the level distance; 
this minimum gap vanishes for $N\to \infty $. In fact,
at $\la =1$ the large $N$ scaling behaviour \cite{ley,dus} is found 
\beq
\Delta E_k\sim \bigg ({k\over N}\bigg )^{1/3}
\label{scal1}
\eeq
whereas for $\la >1$ the gap around $E_{k_c}$ behaves as \cite{ley}
\beq
\Delta E={2\pi \sqrt{\la ^2-1}\over \ln N}.
\label{scal2}
\eeq

We mention that a mean field approach yields the result \cite{gehe}
\bea
E_k(\la )&=&k\, \sqrt{1-\la ^2}\quad {\rm for} \quad \la<1 \\
E_k(\la )&=&k\, \sqrt{2(\la ^2-1)}\quad {\rm for} \quad \la>1 \label{asy}
\eea
being obviously unable to reproduce the richer structure described above; however,
the last two equations do describe the lower part of the spectrum remarkably well.

In the present note we aim at a more complete description of the thermodynamic limit
of the Lipkin model.

\section{The Spectrum} A closer look at the essential features of the spectrum for
large $N$ may elucidate statements made in the previous section and provides clues
of what can be expected in the limit $N\to \infty $. In Fig.1 the spectra are drawn
for a few values of $\la $. As this is done for large values of $N$ we use the
'continuous' variable $x=2k/N$ ranging from zero to unity and plot 
$\epsilon (x)=2 E_k/N$ {\it versus} $x$. Using this scale we note the following
features: for $\la \ge 1$ each spectrum has a singularity when it crosses the critical
line $\epsilon =-1$. In fact the singularity is a point of inflection with a zero
derivative and an infinite second derivative as shown in Fig.2 and in accordance with 
Eq.(\ref{scal2}) (for $\la =1$ the spectrum just touches the critical line with zero
first and infinite second derivative). 
In other words, the level density is singular at the critical line.
Note that the singularity occurring at the critical line is common to all spectra 
for $\la >1$ irrespective of the particular value for $\la $. For $\la <1$ the 
spectra, i.e.~their level densities, are smooth.

\begin{figure}[t]
\begin{center}
	\epsfig{file=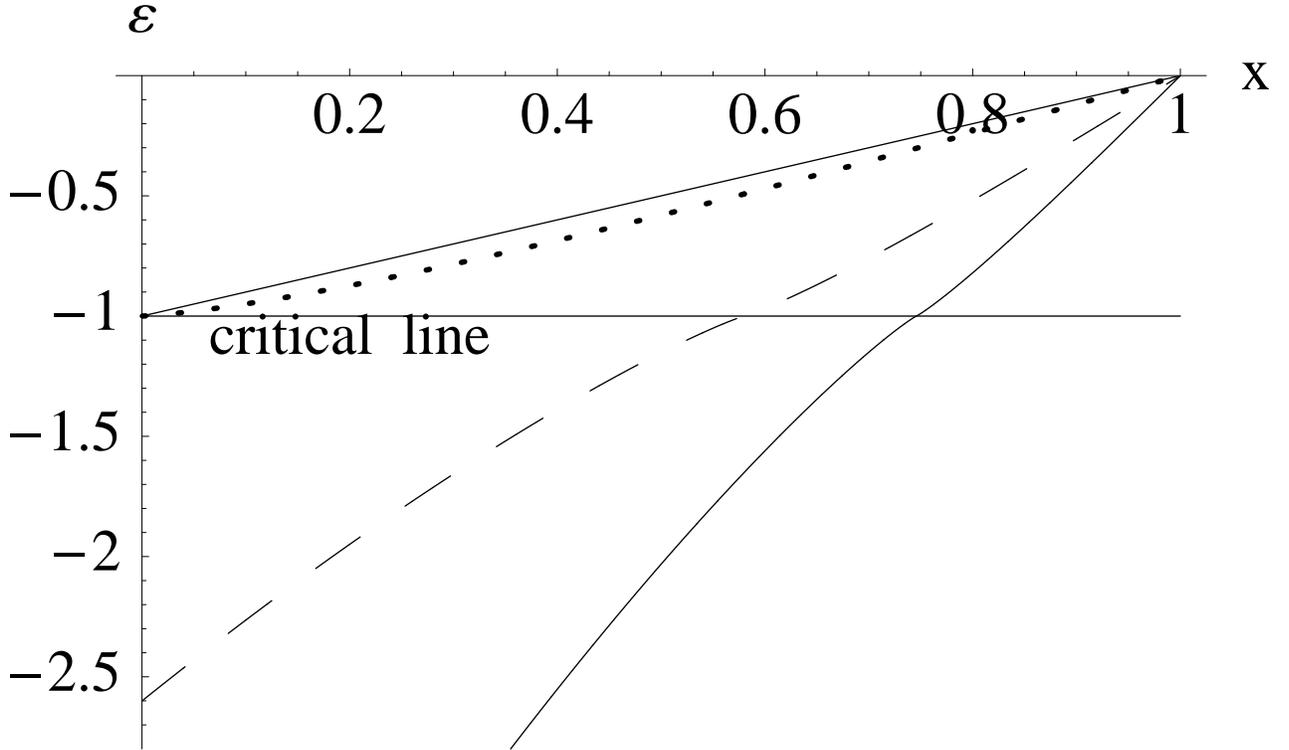,height=10.0cm,clip=,angle=0}
	\caption{Spectra for $\la =0$ (solid line), $\la =1$ (dotted line),
$\la =5$ (dashed line) and $\la =10$ (solid line).}
	\label{spec}
\end{center}
\end{figure}

The range $0>\epsilon >-1$ is the
normal phase while $\epsilon <-1$ corresponds to the deformed phase. For better
illustration we do not display the lower end of Fig.1, it suffices to note
that \cite{ley} for $\la >1$ the ground state assumes the value 
\beq
\epsilon(0)=-{1\over 2}(\la +{1\over \la }). 
\label{gr}
\eeq
The critical line is crossed by each level sequence at
a certain $x_c(\la )$ associated with $\la >1$; the larger $\la $ the larger $x_c$,
i.e.~the phase transition occurs higher up in the spectrum and higher above the ground 
state given in Eq.(\ref{gr}). The critical line is special 
not only as it signals the phase transition through the singularity, but it is also
associated with a specific state vector being identical to the unperturbed ground state
once the limit $N\to \infty $ is attained.
In fact, it is
\beq
{2\over N}H(\la )|j,-j\rangle =-|j,-j\rangle +\sqrt{2}{\la \over N}|j,-j+2\rangle 
\label{state}
\eeq
where the second term of the right hand side vanishes in the limit.
The right hand side also becomes independent of $\la $ when $N\to \infty $. 
Recall that for finite $N$
all levels and state vectors are analytic functions of $\la $ and analytically connected
in the $\la $-plane. As a special result we further note that this state is perfectly  
localised; all other states are more or less extended superpositions of vectors
$|j,m\rangle,\; -j\le m\le j$; this localisation was noted also in \cite{ley}.

\begin{figure}[t]
\begin{center}
	\epsfig{file=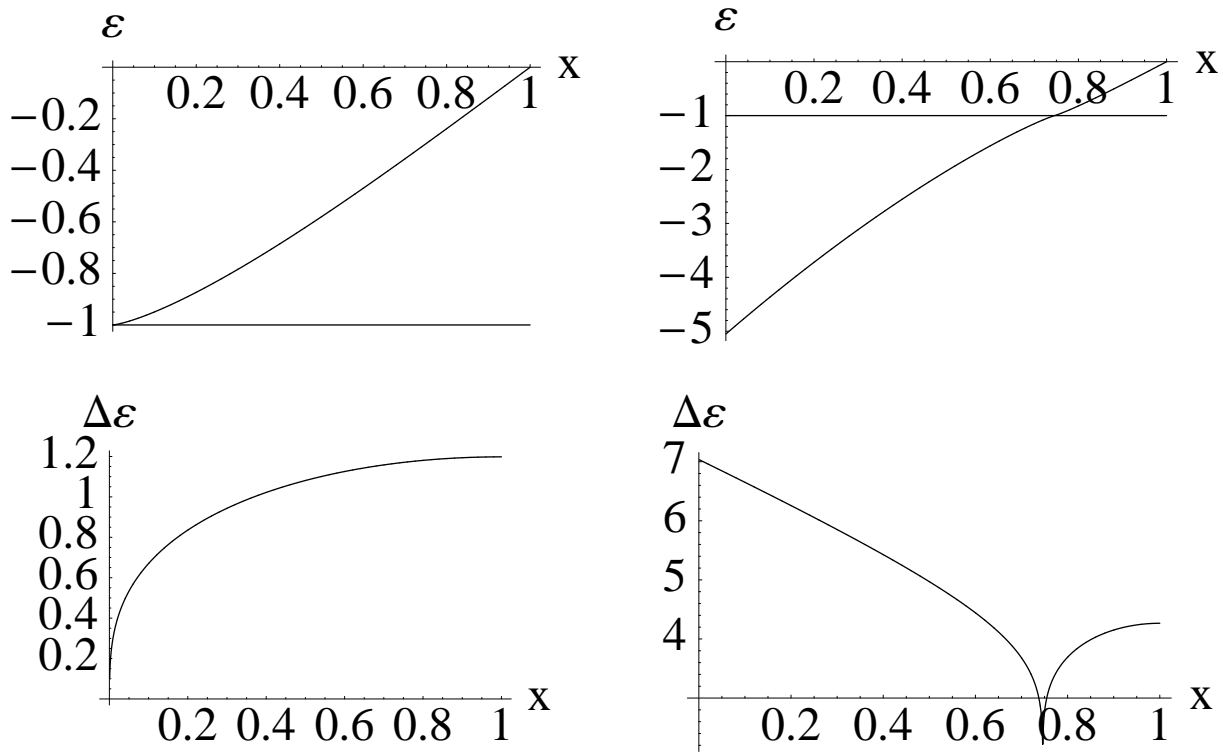,height=10.0cm,clip=,angle=0}
	\caption{Spectra (top row) and their derivatives (bottom row) for
$\la =1$ (left column) and $\la =10$ (right column). The top row is displayed
for convenience, it shows the same respective curves including the critical line as
those in Fig.1.}
	\label{spec}
\end{center}
\end{figure}

The singularity discussed above for the level density is found as well when a particular
level (fixed value of $x$) is plotted {\it versus} $\la $. The same pattern prevails:
there is a point of inflection at $\la =\la _c$ (associated with $\epsilon (x_c)=-1$) with 
a vanishing derivative and an infinite second derivative in $\la $.

\section{Exceptional Points} The whole pattern, in particular the singular behaviour
at the critical line, i.e.~at the phase transition, can be understood from the global
analytical structure of the spectrum as a function of the complex parameter $\la $.
The exceptional points (EP) \cite{kato} are the only singularities of the spectrum
for the underlying model. For a hermitian problem they can occur only in the complex
$\la $-plane. They are square root branch points where, generically, two energies are
connected by a branch point. For the Hamiltonian (\ref{lip}) all energies $E_k(\la )$
are analytically connected, in fact, each $E_k(\la )$ represents the values on a
particular Riemann sheet of one analytic function \cite{hesa}.

\begin{figure}[t]
\begin{center}
	\epsfig{file=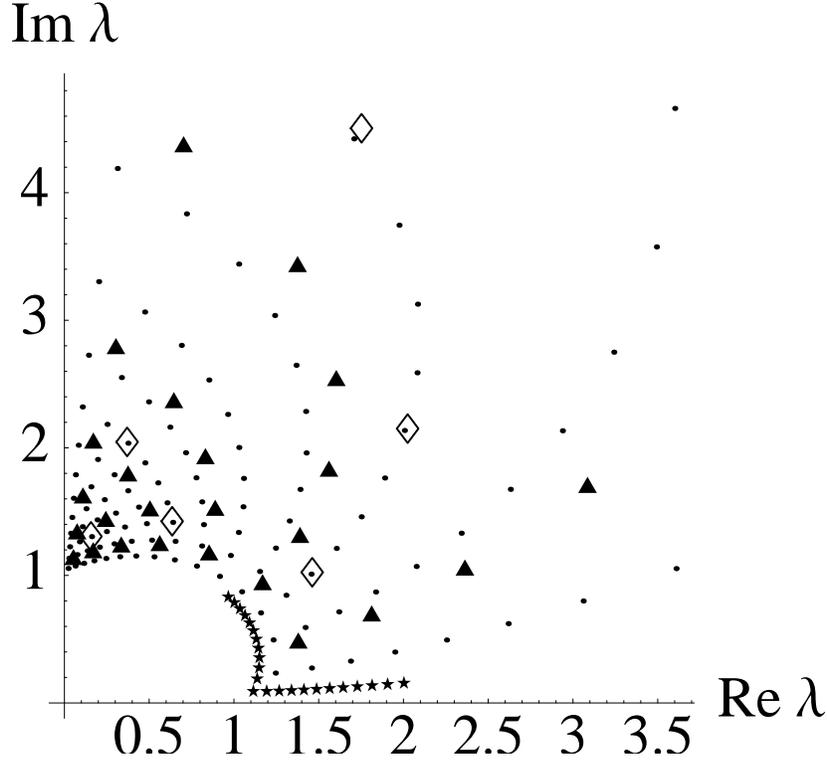,height=10.0cm,clip=,angle=0}
	\caption{Exceptional points in the $\la $-plane for $N=8$ (diamonds), 
$N=16$ (triangles) and $N=32$ (dots).
The first twelve points along the real axis and the innermost arc running 
from the real to the imaginary axis are indicated by asterisks for $N=96$.}
	\label{exc}
\end{center}
\end{figure}

In Fig.3 we illustrate the exceptional points for a few small values of $N$. Two
aspects are of importance: (i) for increasing $N$ some EPs accumulate along the real axis
$\la \ge 1$ and, (ii) at the EPs of the sequence along the real axis, starting at $\la =1$,
the levels coalescing are $E_1$ and $E_2$, $E_2$ and $E_3$, $E_3$ and $E_4$ and so forth.

The second point clarifies immediately some observations made in the previous section.
Take, for instance, $\la =2$. From Fig.3 we read off that there are $N/8$ EPs along the real
axis between $\la =1$ and $\la =2$. This implies that the Riemann sheet (energy level)
with $k=N/8$, that is for $x=0.25$, has a singularity at $\la =2$. In general terms,
there is just one $x_c$ associated with a particular value of $\la $.

With regard to the first point we note two aspects that are, however, related. In the limit
$N\to \infty $ there is a dense set of EPs, say at $\la =2$, for $x=0.25$. The singularity,
being an accumulation of square root branch points, cannot be expected to be of a square
root nature. The more likely behaviour is that of a logarithmic singularity. This is
supported by a similar, albeit simpler situation \cite{herot}, where an accumulation
of square root branch points has been demonstrated analytically to lead to a logarithmic
branch point. There is further strong numerical evidence: a perfect fit is obtained of the
curves in Figs.1 and 2 using the trial functions
\beq 
(x-x_c)^2 (a_1\ln|x-x_c|+a_2(\ln|x-x_c|)^2+\ldots )
\label{fit}
\eeq
employing three terms or less. The trial function (\ref{fit}) is used to fit the spectra
of Fig.1 (after a shift upward by one unit to make the critical line occurring at zero). 
In a subsequent acid test the derivative of (\ref{fit}) is compared with
the respective curves for $\Delta \epsilon $ in Fig.2 yielding an excellent agreement.
Note that the modulus of $x-x_c$ is being used implying that the
function is non-analytic at this point. Depending on the sign of $x-x_c$
a separate fit must be made.
As it was mentioned in the previous section the same
pattern is found when a level is considered as a function of $\la $. A good fit is 
obtained replacing $x$ and $x_c$ by $\la $ and $\la _c$, respectively, in (\ref{fit}).

Not only do we have evidence for a very special singularity, the fact that the EPs
accumulate on the real axis indicates that the thermodynamic limit of the Hamiltonian
(\ref{lip}) {\sl cannot be a hermitian operator}. Moreover, if the limit is taken either in
the deformed or in the normal regime, the two regimes become disconnected for
real values of $\la $.

\section{Conclusion} We summarise our findings: the thermodynamic limit $N\to \infty $ of the
Hamiltonian (\ref{lip}) leads to a specific singular behaviour associated with the phase
transition. It is immaterial as to whether a specific level is considered as a function
of $\la $ or whether the whole spectrum is considered for a fixed value of $\la $. In either
case, a logarithmic singularity is found at a particular value of $\la $ being uniquely 
related to a specific level. Note that the singularity is non-analytic in that, on the real
$\la $-axis, the behaviour on the left hand side of the singularity is not the continuation
of that on the right hand side. The singularity produces an infinity in the respective
second derivative. As such, it is no longer a second order phase transition as it is found for
finite $N$.

Furthermore, the fact that the EPs accumulate on the real $\la $-axis in the limit, renders
the operator nonhermitian in the limit. In fact, depending on how the limit is taken, a 
different result emerges: one operator describing the normal phase and the other describing
the deformed phase. In contrast to the case of finite $N$, the two are dissociated. This is
in contrast to finite $N$, where all energies and state vectors are analytically connected.
In particular, when the limit $N\to \infty $ is attained,
 the state vector $|j,-j\rangle $ (see Eq.(\ref{state})) is common to all spectra, 
irrespective of $\la $. As a particular feature, this vector has,
for $\la >0$, a non-zero overlap with any
other vector of either the normal or the deformed sector. This is in contrast to finite
$N$, where (\ref{lip}) yields the usual complete orthonormal basis for any given $\la $.

It is worth to mention that the semiclassical approach \cite{ley} has explicitly given rise
to ambiguities of a hermitian thermodynamic limit. An unambiguous limit could be obtained
only to lowest order in $\hbar \sim 1/N$.

We refrain from speculating about physical consequences of our findings. Rather we feel
that the thermodynamic limit of the Lipkin model is unsuitable to describe a physical
system despite its beauty for finite yet large values of $N$. We see the value of this
analysis, apart from its own sake, in a specific contribution of the general understanding
of models describing quantum phase transitions.

\vskip 1cm

{\bf Acknowledgment} 

The author gratefully acknowledges helpful discussions with Hendrik Geyer and Frederic Scholtz.

\end{document}